# Stochastic fluctuations promote ordered pattern formation of cells in the Notch-Delta signaling pathway

Madeline Galbraith[1,2], Federico Bocci[3], José N. Onuchic[1,2,4,5*]

[1]Center for Theoretical Biological Physics, Rice University, Houston, TX 77005, USA
[2]Department of Physics and Astronomy, Rice University, Houston, TX 77005, USA
[3]NSF-Simons Center for Multiscale Cell Fate research, University of California Irvine, CA 92697, USA
[4]Department of Chemistry, Rice University, Houston, TX 77005, USA
[5]Department of Biosciences, Rice University, Houston, TX 77005, USA
(Received Feb 1 2022)
*author to whom correspondence should be addressed: jonuchic@rice.edu

The Notch-Delta signaling pathway mediates cell differentiation implicated in many regulatory processes including spatiotemporal patterning in tissues by promoting alternate cell fates between neighboring cells. At the multicellular level, this "lateral inhibition" principle leads to checkerboard patterns of Sender and Receiver cells. While it is well known that stochasticity modulates cell fate specification, little is known about how stochastic fluctuations at the cellular level propagate during multicell pattern formation. Here, we model stochastic fluctuations in the Notch-Delta pathway in the presence of two different noise types – shot and white – for a multicell system. The results show that intermediate fluctuations reduce disorder and guide the multicell lattice toward more checkerboard like patterns. By further analyzing cell fate transition events, we demonstrate that intermediate noise amplitudes provide enough perturbation to facilitate "proofreading" of disordered patterns and cause cells to switch to their correct ordered state. Conversely, high noise can override environmental signals coming from neighboring cells and lead to switching between ordered and disordered patterns. Therefore, in analogy with spin glass systems, intermediate noise levels allow the multicell Notch system to escape frustrated patterns and relax towards the lower energy checkerboard pattern while at large noise levels the system is unable to find this ordered base of attraction.

## I. INTRODUCTION

During many developmental and physiological processes, cells integrate information from their neighbors and the local microenvironment to attain precise patterns in space and time by communicating through several signaling pathways. The Notch pathway is an evolutionarily conserved cell-cell signaling pathway integral to cell differentiation in a range of biological processes including angiogenesis [1], neurogenesis [2], embryogenesis [3], and vein boundary formation of the fruit fly wing [4]. At its core, Notch-Delta signaling operates as a two-cell toggle switch that leads to opposite states between neighboring cells [5]. The signaling activates when the Notch transmembrane receptor binds to the Delta transmembrane ligand of a neighboring cell and the Notch Intracellular Domain (NICD) is cleaved from the Notch receptor [Fig. 1(a)]. Once cleaved, the NICD is transported to the cell nucleus where it upregulates Notch and downregulates Delta [6–8]. This double negative feedback loop leads to opposite fates between a "Sender" cell (high Delta, low Notch) and a "Receiver" cell (low Delta, high Notch) [9]. In addition, if the Notch transmembrane receptor binds to the Delta ligand of the same cell, both Notch and Delta are degraded (cis-inhibition) [10,11].

When generalized to a multicellular scenario, lateral inhibition leads to patterns with alternating cell states, such as in the bristle patterning of the fruit fly [4,12]. The small differences between initial values of Notch and Delta in neighboring cells are amplified by the Notch-Delta negative feedback to generate precise patterns [10,12,13]. While Notch-Delta lateral inhibition is generally accepted as a qualitative model of alternate cell patterning, the precise mechanism by which cells can enforce robust patterning in a noisy cellular environment remains poorly understood. The role of stochasticity in the Notch-Delta pathway is of significant interest to understand the roles of the environment in tissue level patterning. While several theoretical frameworks have been proposed to model Notch-Delta lateral inhibition, multicellular pattern formation has mostly been studied with deterministic models [9,10,12–14], despite the importance of noise in the Notch-Delta pathway in creating stable pattern formation [15–17].

In biochemical signaling networks, stochasticity is often present in the form of either thermal and small number fluctuations during transcription and protein binding (intrinsic noise) or cell-to-cell variability due to external signals in the local microenvironment (extrinsic noise) [18–21]. These sources of stochasticity can be incorporated in models of regulatory dynamics and cell-cell signaling utilizing a system of stochastic differential equations. Different models of fluctuations have been previously employed, including white and shot noise [22]. It has been suggested that transcriptional noise can break the symmetry between initially similar cells with comparable levels of Notch and Delta, thus promoting opposite fates via lateral-inhibition [23]. Further, de Back and collaborators implemented noise in a system containing the Notch pathway, allowing for modulation of the system and symmetry breaking independent of noise source [24]. However, a study of both bistable and tristable gene regulatory networks by Lu and collaborators showed that white and shot noise have different effects on the stability of multistable genetic switches [22]. Lu comments that the real noise in cells may be in an intermediate regime between white and shot noise [22]. Despite these promising initial steps, the implications of white and shot noise on Notch-driven multicell signaling and patterning have not yet been quantified.

Here, we study the spatiotemporal dynamics of a multicellular Notch-Delta lattice model under the effect of stochastic fluctuations. First, we show that white and shot noise have profoundly different effects on the pseudopotential landscape of a single cell interacting with a fixed extracellular environment. While white noise tends to merge the Sender and Receiver states, shot noise more effectively maintains the bistability between states. By generalizing the model to a multicellular scenario, we quantify the robustness of the checkerboard pattern and further show that Notch-Delta signaling maximizes patterning order when operating at an intermediate noise level. While low to intermediate noise supports order by flipping incorrect cell states, high noise can potentially flip correct cell states, thus decreasing pattern ordering. These results suggest an interesting parallel between the mechanism of multicellular pattern formation and the navigation of the energy landscape of spin glass models to better understand how different conditions result in ordered or disordered patterns. Overall, our analysis provides new mechanistic insights into the spatio-temporal patterning driven by Notch-Delta signaling and demonstrates how precise ordering is achieved in the noisy physiological environment.

## II. GENERALIZING THE NOTCH-DELTA SWITCH

To study the role of noise in Notch-Delta signaling dynamics, we first consider the simplified case of a single cell that is exposed to fixed levels of Delta ligands ($D_{EXT}$) and Notch receptors ($N_{EXT}$) that bind to cellular Notch and Delta. These external signals represent the effect of neighboring cells in Sender or Receiver states. Therefore, the single cell model can be interpreted as a mean field approximation of a multicellular lattice model.

To build the stochastic single cell model, we generalize the Notch-Delta switch developed by Boareto and collaborators [9] to include stochastic fluctuations. The temporal dynamics of Notch (N), Delta (D) and Notch Intracellular domain (NICD or I) copy numbers in a cell is modeled with coupled stochastic differential equations:

$$\frac{dN}{dt} = N_0 H^S(I, I_0, n_I, \lambda_N) - k_c ND - k_t ND_{EXT} - \gamma N + G_N(\sigma) \, dW(t), \quad (1)$$

$$\frac{dD}{dt} = D_0 H^S(I, I_0, n_I, \lambda_D) - k_c DN - k_t DN_{EXT} - \gamma D + G_D(\sigma) \, dW(t), \quad (2)$$

and

$$\frac{dI}{dt} = k_t ND_{EXT} - \gamma_I I. \quad (3)$$

The model includes protein production, degradation, chemical binding between Notch and Delta, release of NICD, and transcriptional regulation by NICD on Notch and Delta (parameters values are presented in Table S1 and details of the model are expanded upon in Appendix A1). Notch and Delta are produced with basal production rate constants $N_0$ and $D_0$, which are further modulated by the transcriptional activation or inhibition by NICD, respectively. This modulation is modeled with the shifted Hill function:

$$H^S(I, I_0, n, \lambda) = \frac{1 + \lambda(I/I_0)^n}{1 + (I/I_0)^n}. \quad (4)$$

Once the amount of NICD in the cell (I) crosses the threshold of NICD near which the regulation occurs ($I_0$), the shifted Hill function is saturated. This magnitude of upregulation or downregulation is represented by the foldchange ($\lambda$), while the sensitivity to changes in NICD levels is represented by the Hill coefficient n.

Cellular Notch receptors and Delta ligands can bind to other exogenous ligands and receptors ($D_{EXT}$ and $N_{EXT}$) with rate constant $k_t$, leading to NICD release (trans-activation). Binding of Notch and Delta molecules within the same cells occur with a rate constant $k_c$ and typically leads to degradation of the ligand-receptor complex without further downstream effects

(cis-inhibition). Notch, Delta, and NICD also degrade with basal rate constants ($\gamma$ and $\gamma_I$) due to single molecule degradation and dilution. Finally, the rightmost terms in Eqs. (1)-(2) model stochastic fluctuations either via white or shot noise.

In the absence of noise (i.e., the stochastic terms in Eqs. (1) and (2) are omitted), the single cell behaves either as a monostable or bistable switch depending on the levels of $D_{EXT}$ and $N_{EXT}$. The two monostable regions correspond to a Receiver state (high Notch, low Delta) and a Sender state (low Notch, high Delta). Inside the bistable region, the cell either falls into the Receiver or Sender states based on initial conditions. Different ($D_{EXT}, N_{EXT}$) parameter combinations modulate the basin of attraction of the two states resulting in the bistable region having a non-constant probability of attaining Receiver [Fig. 1(b)].

Introducing stochasticity provides insight into the stability of the Receiver and Sender states. Here, we study the response to two types of noise to mimic potential cellular events and microcellular environments: white and shot noise. While white noise intensity is independent of Notch and Delta concentrations, shot noise is proportional to the cellular concentration of Notch and Delta (See Appendix A1 for details).

The addition of noise introduces stochastic fluctuations around the stable fixed points of the deterministic model, which can be visualized in a pseudopotential landscape $U(N, D) = -\log(P(N, D))$, where $P(N, D)$ is the probability to observe levels of Notch and Delta equal to N and D, respectively (see Appendix B1 for details of the pseudopotential implementation). Within the bistable parameter region, noise generates a landscape with two attractors, from which the location of the minima and the barrier height separating them can be quantified [Fig. 1(c)]. As expected, increasing either white or shot noise progressively decreases the height of the barrier separating Sender and Receiver states while increasing their basin of attraction [Figs. 1(d) and S1(a)-(b)]. Increasing white noise, however, progressively brings the two minima of the landscape closer. Conversely, the separation between Sender and Receiver states is maintained, if not increased, when increasing shot noise [Fig. S1(c)]. Therefore, even though the stability of these states decreases for larger noise amplitudes regardless of noise type, the Notch-Delta switch is able to maintain a clear separation between states when exposed to shot noise. Conversely, strong white noise cannot be sustained effectively, leading to a progressive merging of the two states, in good agreement with previous studies of other small gene circuits [25].

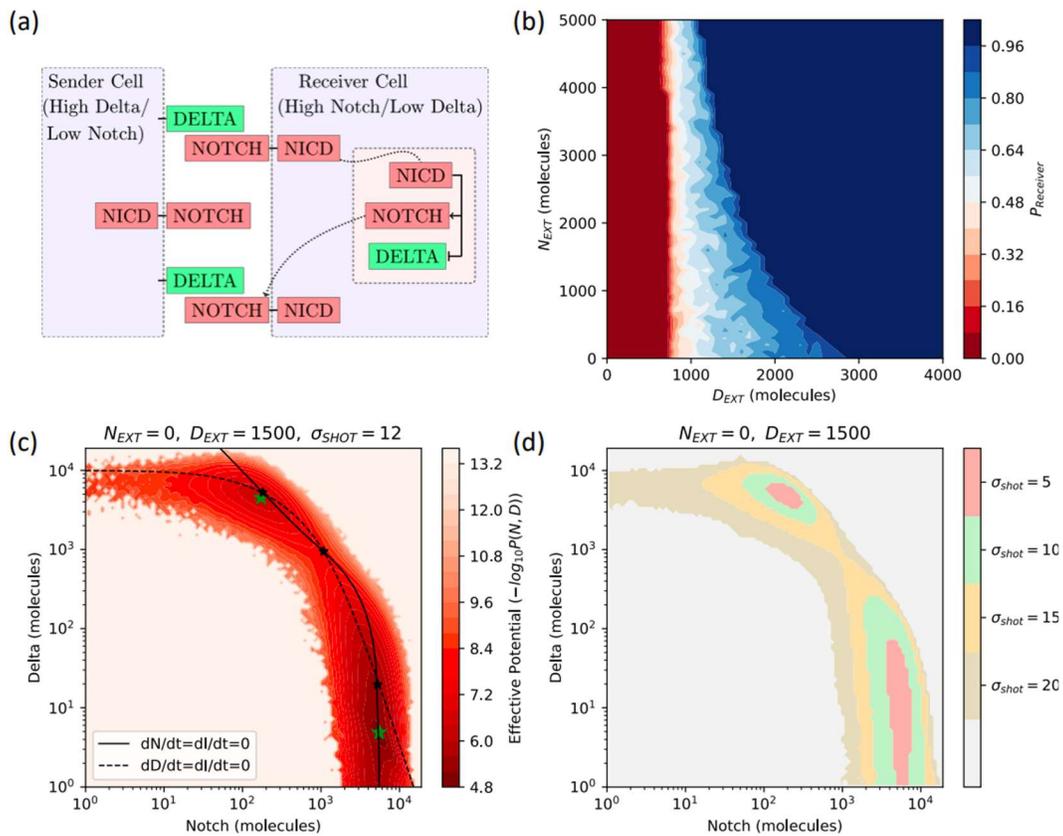

FIG. 1. **Effect of noise in the Notch-Delta single cell model.** **(a)** Schematic of the Notch-Delta circuit. The Delta ligand of the Sender cell binds to the Notch transmembrane receptor of the Receiver cell. Upon binding, the Notch Intracellular Domain (NICD) translocates to the cell nucleus (pink shading), where it upregulates the expression of Notch while inhibiting the expression of Delta. **(b)** Phase diagram of the deterministic single cell model as a function of external Delta ligands ($D_{EXT}$) and Notch receptors ($N_{EXT}$). Color scale indicates the probability to attain a Receiver state. To estimate the Receiver state probability, n=1000 independent trajectories starting from randomized initial conditions were solved for each ($N_{EXT}$, $D_{EXT}$) combination. In the white region where the red shifts to blue, the system is bistable and can be in either the Sender or Receiver state. **(c)** Pseudopotential landscape for a case of shot noise ($\sigma_{shot}=12$, $N_{EXT}=0$, $D_{EXT}=1500$). Continuous and dashed black lines depict the nullclines of the corresponding deterministic model, while black starred dots highlight the stable fixed points of the corresponding deterministic model. Green starred points locate the minima of the landscape. **(d)** Heatmap showing the extension of the Sender and Receiver minima as a function of shot noise for ($N_{EXT}=0$, $D_{EXT}=1500$). The areas of the two states are defined as the regions where the pseudopotential is increased by at most a unit from its value in the minimum.

## III. MODELING THE MULTI-CELL NOTCH-DELTA LANDSCAPE

### A. A deterministic Notch-Delta multicell model leads to patterning disorder

The single cell model provided insight on the robustness of Sender and Receiver states in the presence of a noisy environment. To elucidate the effect of noise on pattern formation in a multicellular environment, we generalize the lateral inhibition mechanism to a two-dimensional multicell scenario. In the simplest case, a two-dimensional square lattice, the lateral inhibition mechanism leads to a stable checkerboard pattern with alternating Senders and Receivers [4,12]. Equations (1)-(3), which model the behavior of a single cell, are extended to account for the binding of exogenous ligands and receptors to cellular ligands and receptors by defining $N_{EXT}$ and $D_{EXT}$ as the average levels of Notch and Delta in the nearest neighbors (details in Appendix A2). Given the interaction between cells and our interest in how individual cellular fluctuations affect order, a lattice size must be chosen that ensure the system is not highly connected. For example, a small lattice with a length of four cells has a short timescale for both cellular and lattice changes. Therefore, we choose a lattice of length sixteen cells where cellular and lattice changes occur at different timescales. Larger lattices would require a longer timescale for pattern relaxation but their general properties should not be dependent on lattice size.

Before studying the temporal and spatial dynamics of pattern formation in a noisy environment, we first establish the deterministic properties of the model. In the absence of noise ($\sigma=0$), a cell layer starting from randomized initial levels of Notch, Delta, and NICD relaxes to a frustrated pattern with many "incorrect" contacts (Sender/Sender and Receiver/Receiver) [Fig. 2(a) and movie M1]. Interestingly, an initial pattern with only one Sender surrounded by Receivers (or "nucleating" case) causes a spatio-temporal cascade that results in a perfect checkerboard pattern [Fig. 2(b) and movie M2]. Thus suggesting, in deterministic cases, only certain initial conditions allow the system to evolve to a checkerboard pattern.

To quantify the deviation from checkerboard patterning, we calculate the percent of "correct" (Sender/Receiver) contacts. Thus, 100% of correct contacts represents a perfect checkerboard pattern, a random pattern would have about 50% of correct contacts, and a pattern where all cells have the same fate would have no correct contacts (See Appendix B 1-2 for details). Analyzing the dynamics of the "nucleating" initial condition shows the system consistently increases in order and reaches a perfect checkerboard pattern on a timescale of about one hundred hours, qualitatively consistent with Notch-driven multicell patterning in multiple developmental systems [5]. Conversely, the randomized initial condition increases in order but leads to a frustrated pattern with only ~75% correct contacts [Fig. 2(c)], further indicating the Notch-Delta mediated multicell systems are predisposed to order but can become trapped in frustrated systems.

To test whether "incorrect" contacts (Sender/Sender or Receiver/Receiver) were propagated during pattern formation, we set up several cell layers with very specific initial conditions. These initial conditions include (1) one quadrant of Senders with the rest Receivers,

(2) one quadrant of Receivers with the rest Senders, (3) one row of Receivers with the rest Senders, (4) one row of Senders with rest the Receivers, and (5) the top half Receivers and the bottom half Senders (Fig. S2). None of the lattices reached a checkerboard pattern upon equilibration, indicating that the multicell Notch-Delta switch can easily remain "stuck" in configurations with "incorrect" contacts between Senders and Receivers, reminiscent of patterning mistakes in spin systems.

We further systematically analyzed the stability of the checkerboard pattern in the noise-free limit, by studying its response to different perturbations. First, discrete perturbations in a checkerboard pattern were created by randomly selecting a fixed fraction of cells in a checkerboard pattern and altering their cellular levels of Notch, Delta and NICD, resulting in an initial condition with a percentage of mistakes (details in Appendix A2). If there are less than a quarter of mistakes in the lattice, then the checkerboard pattern can be easily recovered. A higher percentage of mistakes, however, results in a disordered final pattern that deviates further from the checkerboard pattern as more mistakes are introduced in the initial lattice [Figs. 2(d) and S3].

Furthermore, a continuous perturbation was tested by adding Gaussian noise to a perfect checkerboard initial condition. Specifically, each cell in a checkerboard lattice has a Gaussian random variable with mean, $\mu=0$, and standard deviation, B, added to the cellular values of Notch, Delta, and NICD (details in Appendix A2). Once the amplitude of the continuous Gaussian perturbation to the initial checkerboard pattern is comparable to the magnitude of the Notch and Delta copy numbers (B=1000), the final patterns begin exhibiting disorder [Figs. 2(e) and S4].

Thus, in the noise-free limit, the spatial cell distribution must either be initially very similar to the target checkboard pattern or exhibit a very specific initial pattern (e.g., nucleating) to be able to recover the checkerboard, while larger deviations lead to frustrated patterns. Comparable to spin glass energy landscapes, there seem to be a multitude of basins representing both ordered (checkerboard) and disordered/frustrated patterns. Further, the robustness of the checkerboard pattern to small changes suggests the checkerboard basin is surrounded by basins representing increasingly disordered systems as the distance from the checkerboard basin increases.

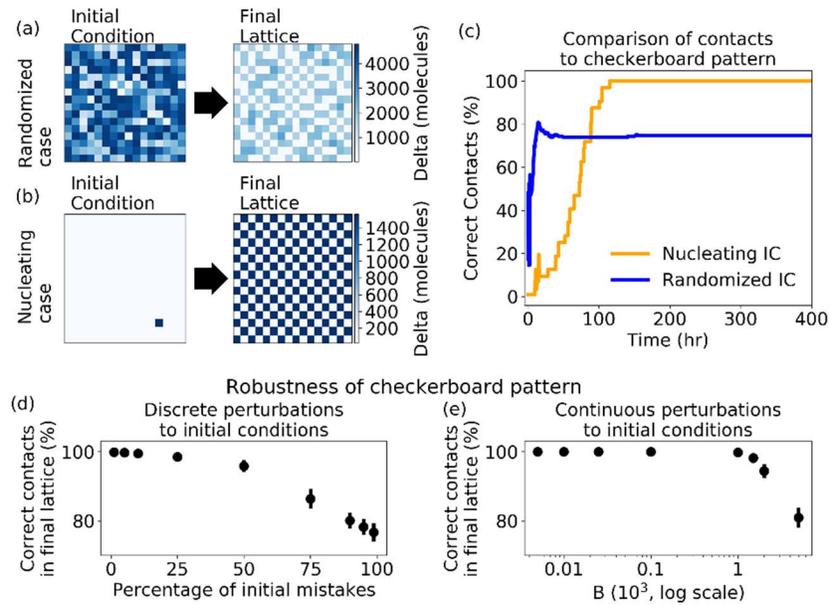

FIG. 2. **Patterning disorder in the noise-free multicell model. (a)** the initial (left) and final (right) patterns starting from randomized initial conditions. The blue heatmap quantifies the cellular levels of Delta. **(b)** Same as (a) for a "nucleating" initial condition. **(c)** The percent of correct contacts as a function of time for the randomized initial conditions (blue) and the nucleating initial condition (orange) corresponding to panels (a) and(b). The lattice with randomized initial condition leads to a frustrated pattern. **(d)** The average percent of correct contacts as a function of percentage of mistakes in the initial checkerboard pattern. **(e)** The average percent of correct contacts as a function of B, the amplitude of the lattice perturbation normalized by $10^3$ (the magnitude of Notch and Delta copy numbers). For panels (d)-(e), percentages of correct contacts are calculated upon full equilibration and averaged over 20 independent simulations. The checkerboard pattern is somewhat robust, allowing for up to a quarter of mistakes in the lattice or a perturbation of magnitude $10^3$, comparable to the Notch and Delta copy numbers.

## B. Optimal noise maximizes lateral inhibition ordering

To elucidate the role of noise in Notch-driven spatial patterning, we generalized the multicell model to include the effect of white and shot noise with stochastic differential equations (see Appendix A1 and A2). Similar to the single cell model, we defined a pseudopotential $U = -\log_{10} P(N, D)$ to identify Sender and Receiver cells. We consider a cell to be in the Sender attraction basin when its pseudopotential energy does not exceed a fixed threshold, chosen to be at a 10-fold difference to the Sender minimum on the pseudopotential landscape (Fig. S5). Likewise, for the Receiver, the cell must be within the threshold of the Receiver attraction basin.

First, we study the dynamics of the correct contacts in the initially randomized lattice as the amplitude of stochastic fluctuations increase (details in Appendix B 2). Near the start of the simulation ([1-10] hr), the system undergoes many drastic changes and quickly relaxes towards a more ordered pattern with around 60%-70% of correct contacts [Fig. 3(a) for white noise and Fig. 3(b) for shot noise]. This is followed by a slower and jumpier equilibration process that separates the systems into distinct levels of correct contacts based on noise amplitude – zero, low, intermediate, or high noise. Notably, the patterning order for very low or very high noise levels is similar to the deterministic model, whereas intermediate noise leads to substantially higher order [Figs. 3(a) and 3(b)]. Thus, the many jumps for the first few simulation hours followed by fluctuations around a single value suggests two timescales for final patterning. During the first stage, the system quickly relaxes towards a more ordered pattern on a timescale associated with Notch signaling equilibration and cell cycle which typically lie within the [10-100] hr interval. The second stage corresponds to fluctuations within the environment that modulate the pattern on a much slower timescale ([100-1000] hr). The effect of stochastic fluctuations can be further decoupled from chemical kinetics with simulations starting from a perfect checkerboard pattern. In this case, noise is the only source for disruption of the pattern; therefore, the fraction of correct contacts progressively separate based on noise level (Fig. S6).

A potential drawback in quantifying disorder based on Receiver-Sender contacts is the ambiguity in defining these cell states in the presence of high noise, as cellular Delta levels do not clearly separate between low (i.e., Receiver) and high (i.e., Sender) [Figs. S7 (a)-(c)]. For this reason, we further define a 'similarity' patterning metric based solely on correlations of Delta levels between neighbors. When applied to the deterministic multicell trajectories of Fig. 2(c), the similarity yields analogous results, thus showing the robustness of our analysis [Fig. S7 (d)].

Comparing similarities at different noise amplitudes confirms the progressive separation of ordered and disordered systems based on noise amplitude, as the level of Delta is modulated even as the pattern is unchanged, allowing for a more apparent separation of order (Fig. S8). Additionally, in agreement with both the correct contact fraction and similarity, the time averaged correlation (details in Appendix B 3) between the final lattice and lattices throughout the simulation shows the two distinct timescales corresponding with the fast chemical kinetics of the system and the subsequent relaxation guided by the stochastic fluctuations, demonstrating that noise corrects disorder (Fig. S9).

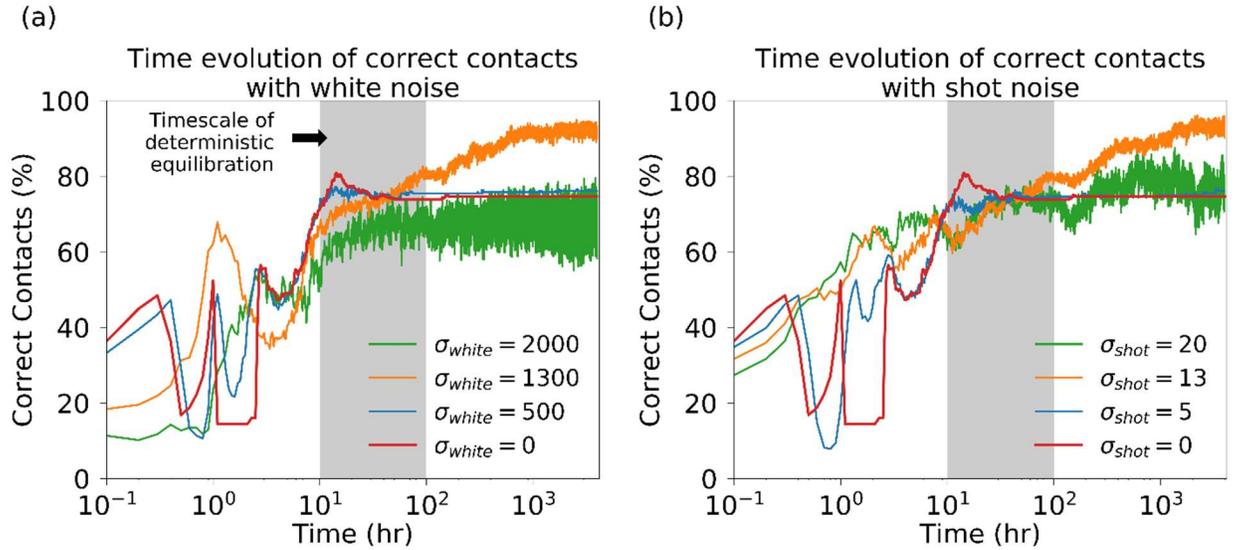

FIG. 3. **Time evolution of pattern formation in the multicellular system.** (a) The percent of correct contacts over time for individual simulations with white noise amplitudes ($\sigma_{white}$=0, 500, 1300, and 2000). The gray region shows the typical timescale of Notch equilibration and cell cycle. The systems start from randomized initial conditions and relax towards more ordered patterns within the first $10$-$10^2$ hr (i.e., Notch equilibration). Afterwards, the stochastic fluctuations dominate, leading to a separation in the order of systems based on noise amplitude. (b) Same as (a) but for corresponding shot noise levels ($\sigma_{shot}$=0, 5, 13, and 20).

While individual simulations provide insight into the timescales of relaxation, we look at aggregated simulations to broadly understand pattern formation. As with the pseudopotential landscape of the one cell system, the basins of attraction are modulated by noise amplitude. When increasing the amplitude of shot noise, a clear separation is maintained between the Sender and Receiver minima [Figs. 4(a)-(c)]. Conversely, the states begin to slowly merge when increasing white noise (Fig. S5), thus confirming our model has consistent response to noise regardless of cell amount (one or many cells).

We further analyzed the aggregated data by comparing the statistics of correct contacts at steady state when starting from either randomized or checkerboard initial conditions for varying levels of noise. This analysis highlights three distinct regimes. In the low noise regime, the randomized lattice exhibits disordered patterns qualitatively similar to the zero-noise limit case (~25% of incorrect contacts), whereas the checkerboard system maintains the checkerboard pattern with nearly 100% of correct contacts [left regions of Figs. 4(d) and 4(e)]. Therefore, low noise levels facilitate sparse exploration of the patterning landscape. In the intermediate noise regime (approximately $900 < \sigma_{white} < 1600$ and $9 < \sigma_{shot} < 16$), the patterning order of the randomized lattices increases and peaks around 92% and 95% of correct contacts (for $\sigma_{shot} = 13$ and $\sigma_{white} = 1300$, respectively) before becoming more disordered at higher noise levels [middle and left regions of Figs. 4(d) and 4(e)]. Further, the difference between the peak in correct contacts for systems starting from randomized initial conditions (i.e., the most ordered system for the shot and white noise cases) is equivalent when also accounting for the standard deviation across all simulations. Conversely, the initially checkerboard pattern becomes less ordered and progressively similar to the initially randomized system as noise increases [middle regions of Figs. 4(d) and 4(e)]. This peak in order implies that, at intermediate noise levels, the system is able to escape the local, frustrated minima, explore the landscape and find ordered patterns that are nearly checkerboard. Finally, in the high noise regime, both randomized and checkerboard initial conditions decrease their order as a function of noise amplitude, suggesting the systems switch too quickly between different basins to relax into a more ordered pattern [right regions of Figs. 4(d) and 4(e)].

The robustness of these trends is further confirmed when inspecting the variation of similarity and cell fractions. First, the similarity metric mimics the trend exhibited by the correct contacts as a function of white or shot noise (Fig. S10). Moreover, a one-to-one ratio of Sender and Receiver cells is maintained throughout most of the noise amplitudes (Fig. S11 for simulations with randomized initial conditions and Fig. S12 for simulations starting from initially checkerboard systems), confirming the systems have the proper ratio to become fully checkerboard.

To further demonstrate the role of intermediate noise in guiding patterning, we test how noise modifies the relaxation of checkerboard patterns with discrete or continuous perturbations. Checkerboard lattices with discrete perturbations [similar to Fig. 2(d)] follow a trend analogous to the randomized initial lattice where the systems are most ordered in the intermediate regime (Fig. S13). Interestingly, in the presence of stochastic fluctuations, checkerboard lattices with continuous perturbations to all cells [similar to Fig. 2(e)] behave comparably to the checkerboard initial condition when perturbations are small and correspond to the randomized initial conditions when perturbations are larger (Fig. S14). Further, the response to noise is consistently observed across lattice sizes, with the trend being more robust as the size of the lattice increases (Fig. S15). These results provide further evidence that both the initial condition (checkerboard vs random/frustrated) and the environment (level of noise) contribute to pattern formation.

The trend of intermediate noise having the most beneficial effect on order is further demonstrated by the distribution of correct contacts explored by the lattice during the simulation (Fig. S16). There is little exploration of the landscape at low noise levels, while intermediate noise levels allow exploration of many configurations with high correct contact fractions, further supporting the analogy that the checkerboard pattern resides in the lowest energy minima and is surrounded by many shallow minima. Conversely, at high noise levels the systems explore a wide range of configurations but are not able to attain ordered patterns.

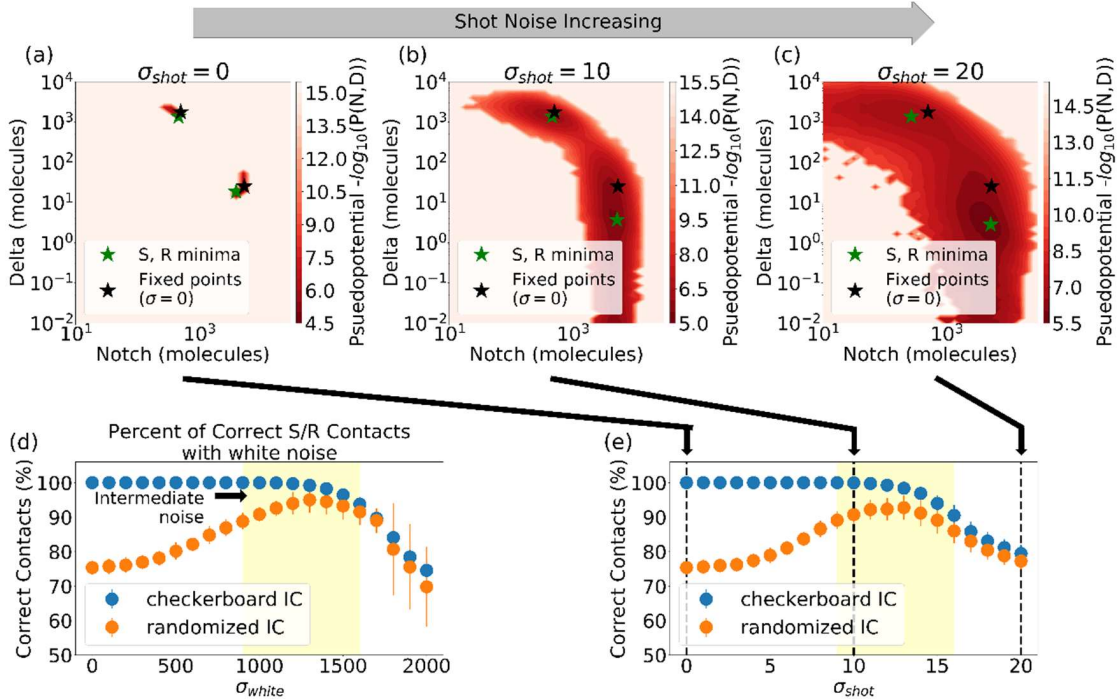

FIG. 4. **Stochastic influence on lateral inhibition and multicell pattern formation. (a)** Pseudopotential landscape for the multicell model in the absence of noise ($\sigma_{shot}=0$) for a single independent simulation. X- and y-coordinates represent cellular levels of Notch and Delta, respectively. Green starred dots depict the location of the landscape minima corresponding to S and R states, while black starred dots depict the location of the S and R states in a perfect checkerboard pattern. **(b)** Same as (a) but for $\sigma_{shot}=10$. **(c)** Same as (a) but for $\sigma_{shot}=20$. Panels (a)-(c) show that shot noise maintains separation of the Sender and Receiver states even as the system explores the state space. **(d)** The percentage of correct contacts (i.e., Sender-Receiver) as a function of white noise for random initial conditions (orange). **(e)** Same as (d) but for correct contacts as a function of shot noise. The lattice is influenced towards a more ordered system for intermediate levels of noise. For panels (d) and(e), results are averaged over 20 independent simulations of 9000 hr using random initial conditions after a 1000 hr relaxation period. For initially checkerboard systems, 20 independent simulations of 4000 hr are averaged after allowing for 1000 hr of relaxation.

### C. Intermediate noise maximizes order by selectively flipping incorrect cell states

To elucidate how patterning and error correction operate at different noise levels, we studied more systematically the statistics of cell switching and its dependence on the local cell environment. To define single cell transitions in a noisy multicell system, we take advantage of the pseudopotential landscape. To complete a transition, a cell must not only exit the attraction basin of its current state, but also enter the attraction basin of the opposite state (Appendix B 4). Following a parallel with the single cell model, a Receiver cell surrounded by a "Sender-like" environment should have a low switching rate, whereas a Receiver cell surrounded by a "Receiver-like" environment should have a higher switching rate.

To quantify the switching as a function of the cell's local environment, we estimate the transition waiting times between Sender and Receiver states as a function of external Notch and Delta defined as the average over the cell's nearest neighbors. In other words, the transition waiting time represents the time spent in a state (Sender or Receiver) before switching to the opposite state. At intermediate noise levels, the Receiver state is stable ($10^3$-$10^4$ hr) when surrounded by a "Sender-like" (high Delta, low Notch) environment, while switching occurs on much shorter timescales (1-10 hr, comparable or shorter to the timescale of Notch equilibration) when surrounded by a "Receiver-like" (low Delta, high Notch) environment [Figs. 5(a) and 5(b)]. Consistently, the Sender state follows the opposite trend, being stable when the environment behaves as a Receiver and unstable when the environment behaves as a Sender (Fig. S17). The stability of cells in the "correct" environment suggests highly ordered patterns lie within deep minima on the landscape while disordered patterns are represented by shallow minima. Strikingly, the connection between external signal (Sender-like vs Receiver-like environment) and transition time is weaker for cases of very high noise ($\sigma_{white}=2000$, $\sigma_{shot}=20$) due to the destabilization of the checkerboard pattern. At these high noise amplitudes, fast transitions were observed when cells occupied the "correct" and "incorrect" states, suggesting stochastic fluctuations are larger than the energy barriers separating ordered and frustrated configurations [Figs. 5(c) and 5(d)]. Therefore, intermediate noise levels provide enough perturbation to "flip" cells that occupy an incorrect state, whereas a very strong noise also enforce transitions in cells that occupy the correct state, thus decreasing the overall ordering. This trend is consistently observed when comparing transition times as a function of Sender/Receiver neighbors instead of average external Delta/Notch. The greatest stability exists for cells in the "correct" environment (Figs. S18 and S19).

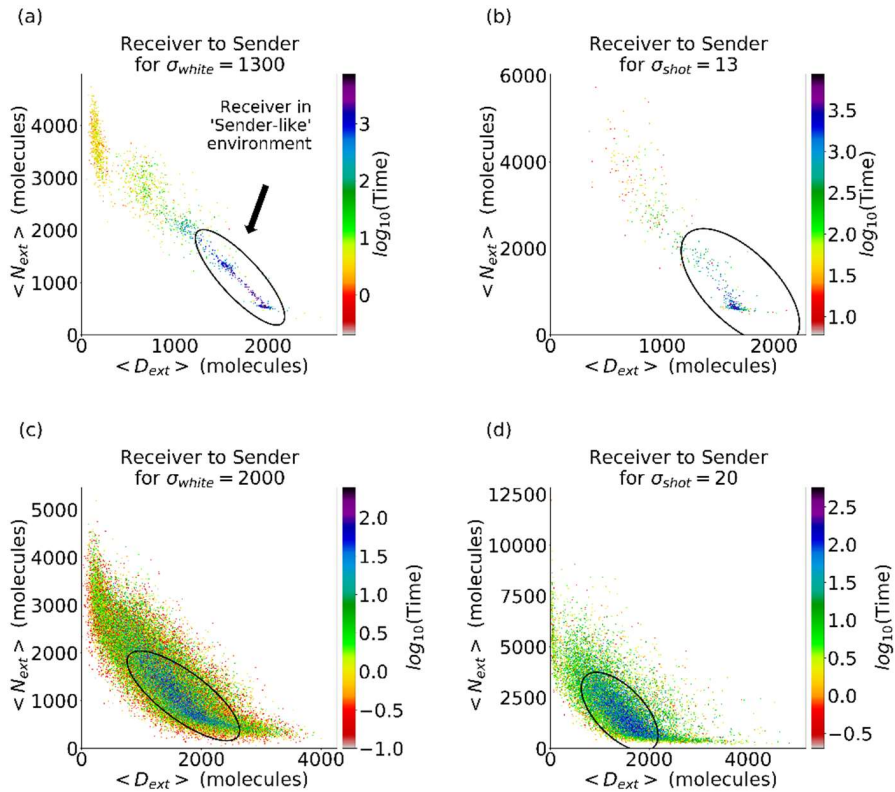

FIG. 5. **Timescales for single cell and lattice transitions. (a)** The distribution of transition waiting times from Receiver to Sender as a function of $N_{EXT}$ and $D_{EXT}$ in a simulation starting from randomized initial conditions with white noise amplitude of $\sigma_{white} = 1300$ for 9000 hr after a 1000 hr relaxation period. Each dot represents the average switching event, while the x- and y-coordinates represent the average levels of Delta and Notch in neighboring cells during the transition. The color scale depicts the waiting time before the transition event. The "correct" state region (a Receiver surrounded by Sender-like environment) is circled, and dots within this region exhibit the longest transition waiting time. **(b)** Same as (a) but for $\sigma_{shot} = 13$. **(c)** Same as (a) but at a high white noise level ($\sigma_{white}=2000$). These larger fluctuations also flip cells from the "correct" to "incorrect" state. **(d)** Same as (a) but for high shot noise ($\sigma_{shot}=20$).

## IV. DISCUSSION

Notch-Delta signaling plays a ubiquitous role in multicellular pattern formation in different physiological and pathological contexts by promoting lateral inhibition between Sender cells (low Notch, high Delta) and Receiver cells (high Notch, low Delta) [7,26]. While noise plays a key role in this signaling [15–18], most of the existing theoretical analysis focuses on deterministic models [9,10,12–14]. Here, we developed a stochastic model of Notch-Delta biochemical signaling to elucidate the role of fluctuations during Notch-driven spatial patterning formation. Our results suggest that appropriate levels of stochastic fluctuations promote an ordered patterning by proofreading the pattern and forcing cells to "flip" to their "correct" state. Our study focused on two types of noise, white and shot, which have been previously shown to capture different regimes of stochastic fluctuations in gene regulatory networks [22].

The connection between multicell patterning and noise was investigated using a two-dimensional model of cells arranged on a square lattice. In the absence of noise, a checkerboard pattern steady state with alternated Sender and Receiver cells could be recovered when the initial pattern was close to the final checkerboard pattern or in the presence of other, specific initial conditions, such as the 'seeding' of a single Sender cell. Randomized initial conditions, however, led to disordered steady states, suggesting the existence of a complex landscape with several local attractors. In the presence of weak stochastic fluctuations, the lattice stays near the comparable zero-noise disordered state; conversely, in presence of high noise the system cannot stabilize the ordered pattern and explores disordered patterns in the configurational landscape. Strikingly, the intermediate noise levels leads to an optimal patterning order. Specifically, analysis of single cell transition trajectories within the lattice demonstrate that intermediate noise can serve as a mechanism for error correction by "flipping" cells to the "correct" state. Our findings that the Notch multicell system can be trapped in a frustrated state if not initially seeded for checkerboard (in the deterministic case) or without the proper level of stochastic fluctuations, agrees with previous results which suggest stochastic fluctuations are critical to correct patterning [15–17,23].

Methods from statistical mechanics, and the Ising model in particular, have been previously used to understand a multitude of biological problems including neural networks, bird flocking, and protein folding [27–30]. In analogy with the energy landscape of conventional spin models, the multi-stability of biochemical and gene regulatory networks can be described by a pseudopotential landscape [31]. This effective energy landscape approach has been previously applied to explain cell fate transitions in other biological systems, such as the Epithelial-Mesenchymal Transition (EMT) in cancer [31]. Qualitatively comparing the antiferromagnetic Ising model, and the associated energy landscape, to the Notch-mediated pattern formation provides insights on how patterns achieve ordering. It has been shown for low temperature spin glasses, that the free energy landscape is very rugged and has many valleys in which the systems become trapped in frustrated states [32,33]. These barriers between valleys block the access of the system to the lowest energy state unless there is an external source modulating the free energy landscape [34].

Following the parallel with a spin glass landscape, our results can be interpreted as follows. In the deterministic case (i.e., zero-temperature limit), Notch-Delta systems can only reach their lowest energy state (i.e., checkerboard) if there is a downhill path they can traverse through the landscape, whereas systems with randomized initial conditions remain trapped in frustrated states. Stochastic fluctuations allow the multicell system to navigate the landscape, overcome energy barriers and access low-energy states. While low noise is sufficient to escape local, frustrated minima and reach slightly more ordered systems, intermediate noise enables a more thorough exploration of the landscape to reach very ordered, low energy states. Fluctuations in the high noise case, however, become larger than the energy barriers separating different stable patterning configurations, causing the system to remain unstable and switch between high and low energy states. Thus, the spin glass energy landscape can provide a useful framework to understand the system-wide effect of noise on Notch-mediated pattern formation.

Given that pattern formation in biological systems is not instantaneous, we considered how the temporal dynamics of individual cells contributed to pattern formation. Our results suggest a short timescale corresponding with the typical equilibration time of the Notch circuit ($10$-$10^2$ hr) where the system attains a rough ordering with many mistakes, and a longer timescale where stochastic fluctuations further 'proofread' the pattern and allow the system to relax towards order. This timescale separation potentially helps explain different patterns observed experimentally. For example, sprouting angiogenesis integrates fate differentiation, proliferation, and migration on timescales comparable with the Notch circuit equilibration (i.e., tens of hours), thus potentially not providing endothelial cells with enough time to navigate the 'patterning landscape'. In this case, lateral inhibition between migrating Tip cells (i.e., Senders) and proliferating Stalk cells (i.e., Receivers) organize the formation of a new blood vessel from the existing lumen [35]. Interestingly, patterning disorder and uneven spacing between sprouts has been recently reported in Human umbilical vein endothelial cells (HUMEC) cultivated *in vitro* [36]. Similarly, the early developmental stages of the avian auditory organ (E8-E9) are characterized by frequent contacts between hair cells (i.e., Senders). This patterning is corrected at later stages, after approximately 4-5 more days, despite unchanged ratios of hair and supporting cells (i.e., Receivers), thus suggesting a role for stochastic fluctuations in cell rearrangement [37].

The current work focuses on the implications of stochastic fluctuations in cell layers mediated by the Notch-Delta pathway. This framework, however, could be generalized in the future to include context-specific aspects of the Notch-Delta pathway and provide quantitative insight on a greater range of biological questions. For instance, epithelial cells often organize in nearly hexagonal geometries that exhibit a lowest energy state where Senders are surrounded by six Receivers [38], as seen for instance in the patterning of hair cells during fruit fly development [26]. Also, the 'standard' Notch-Delta signaling can be modified by additional mechanisms such as long-range interactions via diffusible ligands [39] and variable cell-cell contact area [40]. Moreover, ligands of the Jagged family can compete with Delta and instead lead to lateral induction, or spatial propagation of a similar hybrid Sender/Receiver phenotype that plays a critical role in coordinating collective cell migration during cancer invasion [9,41]. The interplay between Delta and Jagged ligands can regulate critical transitions between lateral inhibition and lateral induction [14]. Finally, Notch regulates, and is in turn sensitive, to mechanical cues in the cellular microenvironment, hence raising interesting questions about the integration of biochemical and mechanical regulation of Notch signaling [42]. Thus, it would be of interest to quantify how cell packing geometry and modifications of Notch signaling modulate the interplay between stochastic fluctuations and patterning. Confirming that optimal noise levels can influence these systems towards the lowest energy state would provide further evidence that noise is critical to Notch-driven patterning.

# AUTHOR CONTRIBUTION

M.G. and F.B. developed the model, performed simulations, and wrote the manuscript. All authors interpreted and discussed the results and edited the manuscript. J.N.O. supervised the research, proposed and discussed the scientific questions and helped with the editing of the manuscript.

# ACKNOWLEDGEMENTS

We thank Peter Wolynes and Dongya Jia for many discussions and suggestion during the development of this work. Research was supported by the National Science Foundation (NSF) Center for Theoretical Biological Physics (NSF PHY-2019745) and the NSF grants PHY 1522550 and CHEM-1614101. J.N.O is a CPRIT Scholar in Cancer Research. M.G. was also supported by the NSF GRFP no. 1842494. F. B. is supported by a grant from the Simons Foundation (grant number: 594598, QN).

# APPENDIX A: SIMULATION DETAILS
## 1. Details of model and simulation for Notch-Delta switch

In this work we begin with the one cell model exposed to a range of external Notch and Delta in the environment ($N_{EXT}$ and $D_{EXT}$, respectively). For each ($N_{EXT}$, $D_{EXT}$) combination the cell relaxes until full equilibration. The stochastic terms added to Eqs. (1) and (2) consist of a noise amplitude multiplied by a Gaussian Normal distribution. In the white noise case, the stochastic term for Eq. (1) is given by,

$$G_N(\sigma_{white}) dW(t) = \sigma_{white} \text{Normal}(0, 1) \, dt \tag{A1}$$

and for Eq. (2) it is given by,

$$G_D(\sigma_{white}) dW(t) = \sigma_{white} \text{Normal}(0, 1) \, dt \tag{A2}$$

In the shot noise case, the functional form for Eq. (1) is

$$G_N(\sigma_{shot})dW(t) = \sigma_{shot}\sqrt{Ndt}\,\text{Normal}(0,1) \tag{A3}$$

while for Eq. (2) it is

$$G_D(\sigma_{shot})dW(t) = \sigma_{shot}\sqrt{Ddt}\,\text{Normal}(0,1) \tag{A4}$$

To understand the Notch-Delta switch modeled by Eqs. (1)-(3), the initial value of N, D, and NICD in the cells was selected and the rate equations were numerically solved using the Euler method with a time step of dt=0.1 hr. The stochasticity of the system is computed by randomly sampling a value from a Gaussian Normal distribution and multiplying it by the corresponding amplitude ($\sigma_{white}$ for white noise or $\sigma_{shot}\sqrt{X}$ for shot where X is either Notch or Delta in the cell). The random variable to be added to Notch, Eq (1), and Delta, Eq (2), are independent samples. Given that Notch and Delta vary on a range of $10^2$-$10^4$ molecules, the amplitude of shot noise is approximately 10-100 times larger than the amplitude of white noise when $\sigma_{white} = \sigma_{shot}$.

## 2. Simulation details of multicell model

The multicell system is a square lattice with periodic boundary conditions to control for cell-cell contact area and cell geometry. Thus, we can assume each cell contributes one quarter of its' Notch and Delta molecules to each of its four neighbors, implemented in our model as

$$N_{EXT} = \frac{1}{4}\sum_{i=1}^{4} N_i(t), \tag{A5}$$

for $N_{EXT}$ in Eq. (2) and

$$D_{EXT} = \frac{1}{4}\sum_{i=1}^{4} D_i(t), \tag{A6}$$

for $D_{EXT}$ in Eqs. (1) and (3).

To simulate the multicell model, the system was set to have a timestep of dt=0.1 hr and solved via the Euler method, as with the one cell model. In the stochastic models, the noise terms were calculated similar to the single cell system by randomly sampling from a Gaussian distribution. Additionally, a different value was sampled for every cell. The deterministic simulations were allowed to equilibrate for 5000 hr to reach a final pattern. The stochastic simulations had a relaxation period of 1000 hr. For each level of noise, a simulation length of 10000 hr, including the 1000 hr of relaxation, was completed for lattices starting from random initial conditions and averaged over 20 independent simulations unless noted. The non-random initial condition lattices were averaged over 20 different simulations with a length of 4000 hr after relaxation.

For a lattice with randomized initial conditions, the value of Notch, Delta and NICD is sampled from a random uniform distribution; the Notch level of the cell in the ith row and jth column is

$$N_{i,j}(t=0) = U(0, 10000), \qquad (A7)$$

the Delta level of the cell in the ith row and jth column is

$$D_{i,j}(t=0) = U(0, 10000), \qquad (A8)$$

and the NICD level of the cell in the ith row and jth column is

$$I_{i,j}(t=0) = U(0, 2000). \qquad (A9)$$

For systems with a specific initial configuration (e.g., checkerboard, nucleating, one quadrant of Receivers, etc.), the lattice is generated using only the values of Notch, Delta, and NICD of the deterministic two-cell model (Table S2). If an initial lattice is said to have a specific percentage of mistakes, then a specific percentage of cells in a checkerboard lattice were randomly selected and perturbed away from a perfect checkerboard. These cells have their values of Notch, Delta, and NICD replaced by randomly sampling a new value from a uniform distribution; the updated value of Notch in the cell is

$$N_{new}(t=0) = U(0, 10000), \quad (A10)$$

the updated value of Delta is

$$D_{new}(t=0) = U(0, 10000), \quad (A11)$$

and the updated value of NICD is

$$I_{new}(t=0) = U(0, 10000). \quad (A12)$$

Lastly, if a checkerboard has been perturbed by a standard deviation, then this lattice was modified from checkerboard. For each cell in the checkerboard lattice, a value sampled from a Gaussian with mean $\mu = 0$, and standard deviation B, was added to the value of Notch, Delta, and NICD (sampled from independent distributions); for Sender cells

$$N_{new}(t=0) = N_{Sender} + BNormal(0, 1), \quad (A13)$$

$$D_{new}(t=0) = D_{Sender} + BNormal(0, 1), \quad (A14)$$

and

$$I_{new}(t=0) = I_{Sender} + BNormal(0, 1). \quad (A15)$$

## APPENDIX B: ANALYSIS METHODS

### 1. Identification of Sender and Receiver states in the multicell model

Spatial constraints and noise give rise to a continuous spectrum of Notch and Delta levels in the multicell model, thus complicating the classification of cells as Sender or Receivers. To determine the state of individual cells and whether the pattern is exactly checkerboard or a checkerboard with a few mistakes, we define the Sender and Receiver states based on the pseudopotential

$$U(N, D) = -\log_{10} P(N, D) \tag{B1}$$

where P(N, D) is the probability of a cell to have a level of N and D molecules of Notch and Delta, respectively. Levels of Notch and Delta in all cells are aggregated to construct the landscape. The two deepest minima are calculated and labeled as either Sender ($\text{Delta}_{minimum} > \text{Notch}_{minimum}$) or Receiver($\text{Delta}_{minimum} < \text{Notch}_{minimum}$). For the Sender, the attraction basin is defined as the region of the landscape where the value of P(N, D) is increased by at most a 10-fold difference from the value in the minimum ($P_{Sender}$). The attraction basin for the Receiver state is defined likewise. A cell that is initially classified as a Sender will switch to the Receiver state if and only if it crosses the threshold to enter the Receiver basin (vice versa for a cell initially classified as Receiver). This condition prevents false positive state switches when large fluctuations transiently displace cells outside of their state thresholds.

## 2. Quantification of pattern disorder

When analyzing the final equilibrated results of the model, we looked at individual cells and the entire lattice collectively. Given that the Sender and Receiver states have Delta values two orders of magnitude different, we developed the similarity metric (S) that defines the distance to checkerboard using only the value of Delta within the cells. The benefit is we do not need to go through the analysis of identifying which state the cell is in while determining how close the pattern is to checkerboard. Defining the similarity metric using

$$S = \frac{1}{4N_{rows}} \sum_{i=1}^{N_{rows}} \left(1 - r(x_i, x_{i+1})\right) + \frac{1}{4N_{cols}} \sum_{j=1}^{N_{cols}} \left(1 - r(x_j, x_{j+1})\right), \tag{B2}$$

where

$$r(x, y) = \frac{\sum_k (x_k - \bar{x})(y_k - \bar{y})}{\sigma_x \sigma_y}, \qquad (B3)$$

allows us to handle the continuous variables and shifting of the Sender/Receiver states. While these definitions work for most cases, if either the row or column has the exact same value of Delta then the similarity metric will not provide an accurate portrayal of the pattern, thus it is important to analyze the final patterns and ensure the similarity is near the expected value (e.g., near S=1 if the pattern looks checkerboard).

We also compute the number of correct contacts in the lattice where a correct contact is defined as two adjacent cells that have opposite fates (S/R or R/S). Each cell in the square lattice can have up to 4 correct contacts with the total number of correct contacts in the lattice maximally 2(NxN) for a square lattice of length N cells for N>= 3. The number of correct contacts for each cell is defined as

$$CC_i = \frac{1}{2} \sum_{nn} \delta(\text{state}_i, \text{Sender}) \delta(\text{state}_{nn}, \text{Receiver}) + \delta(\text{state}_i, \text{Receiver}) \delta(\text{state}_{nn}, \text{Sender}), \qquad (B4)$$

Where the state is determined based on the pseudopotential landscape. If the cell is within the thresholds of the Notch and Delta values of the Sender basin then it is labeled Sender, and similarly for the Receiver state (details in previous section, Appendix B 1).

### 3. Lattice time correlation

The Receiver and Sender states of the Notch-Delta pathway can be transformed to the Up and Down states of the Ising model with a transformation where R=>1 and S=>-1. This transformation is defined using

$$s_i = \begin{cases} 1 & N_i > D_i \\ -1 & D_i > N_i \end{cases} \quad (B5)$$

where the continuous (N, D) variable system is transformed to a $\pm 1$ discrete spin system.

Also, using these transformed states, we can compute the time averaged correlation (q) [43] of each lattice with the initial or final lattice. These two equations can be used to quantify the timescale of similarity between the initial pattern and pattern at a later time

$$q_{initial}(t) = \frac{1}{N} \sum_i^N <s_i(t=0) \, s_i(t)>_T, \quad (B6)$$

and between the final pattern and a pattern at an earlier time

$$q_{final}(t) = \frac{1}{N} \sum_i^N <s_i(t=t_f) \, s_i(t)>_T. \quad (B7)$$

### 4. Calculation of transition waiting times

When calculating the transition time, we use the Sender/Receiver definitions mentioned in Appendix B 1. A successful transition occurs when a cell leaves its current attraction basin (Receive or Sender) and enters the opposite basin of attraction (Sender or Receiver). The transition waiting time begins after a successful transition (Receiver to Sender or Sender to Receiver). It then ends once the reverse transition is successful (Sender to Receiver or Receiver to Sender). The transition waiting time includes both the time spent in a basin of attraction (Sender or Receiver) and the time transitioning between basins (Sender to Receiver or vice versa). This accounts for fluctuations around the thresholds for the attraction basins and reduces the likelihood of misclassifying short or long transition times. Since the cell transitions are tied to the value of Notch and Delta in the neighboring cells ($N_{EXT}$ and $D_{EXT}$), we calculate the transition times at values of $N_{EXT}$ and $D_{EXT}$. To better identify the transitions, and since Notch and Delta are continuous variables, we bin the data by the ranges N = [0,8000] with a step of 10 and D = [0,6000] with a step of 1 and compute the average time for each pair of N and D. Likewise, we can also calculate the transition times as a function of neighbors that are Senders and Receivers.

[1] A. Garcia and J. J. Kandel, *Notch: A Key Regulator of Tumor Angiogenesis and Metastasis.*, Histol Histopathol 27, 151 (2012).

[2] C. C. F. Homem and J. A. Knoblich, *Drosophila Neuroblasts: A Model for Stem Cell Biology*, Development 139, 4297 (2012).

[3] J. Reichrath and S. Reichrath, *Notch Signaling in Embryology and Cancer, Notch Signaling in Embryology*, Adv Exp Med Biol 1218, 9 (2020).

[4] J. F. C. de and S. Bray, *Feed-Back Mechanisms Affecting Notch Activation at the Dorsoventral Boundary in the Drosophila Wing*, Development 124, 3241 (1997).

[5] F. Bocci, J. N. Onuchic, and M. K. Jolly, *Understanding the Principles of Pattern Formation Driven by Notch Signaling by Integrating Experiments and Theoretical Models*, Front Physiol 11, 929 (2020).

[6] W. R. Gordon, D. Vardar-Ulu, S. L'Heureux, T. Ashworth, M. J. Malecki, C. Sanchez-Irizarry, D. G. McArthur, G. Histen, J. L. Mitchell, J. C. Aster, and S. C. Blacklow, *Effects of S1 Cleavage on the Structure, Surface Export, and Signaling Activity of Human Notch1 and Notch2*, Plos One 4, e6613 (2009).

[7] S. J. Bray, *Notch Signalling in Context*, Nat Rev Mol Cell Bio 17, 722 (2016).

[8] M. R. Stupnikov, Y. Yang, M. Mori, J. Lu, and W. V. Cardoso, *Jagged and Delta-like Ligands Control Distinct Events during Airway Progenitor Cell Differentiation*, Elife 8, e50487 (2019).

[9] M. Boareto, M. K. Jolly, M. Lu, J. N. Onuchic, C. Clementi, and E. Ben-Jacob, *Jagged–Delta Asymmetry in Notch Signaling Can Give Rise to a Sender/Receiver Hybrid Phenotype*, Proc National Acad Sci 112, E402 (2015).

[10] D. Sprinzak, A. Lakhanpal, L. LeBon, L. A. Santat, M. E. Fontes, G. A. Anderson, J. Garcia-Ojalvo, and M. B. Elowitz, *Cis-Interactions between Notch and Delta Generate Mutually Exclusive Signalling States*, Nature 465, 86 (2010).

[11] D. del Álamo, H. Rouault, and F. Schweisguth, *Mechanism and Significance of Cis-Inhibition in Notch Signalling*, Curr Biol 21, R40 (2011).

[12] P. Formosa-Jordan and M. Ibañes, *Competition in Notch Signaling with Cis Enriches Cell Fate Decisions*, Plos One 9, e95744 (2014).

[13] J. R. Collier, N. A. M. Monk, P. K. Maini, and J. H. Lewis, *Pattern Formation by Lateral Inhibition with Feedback: A Mathematical Model of Delta-Notch Intercellular Signalling*, J Theor Biol 183, 429 (1996).

[14] T.-Y. Kang, F. Bocci, M. K. Jolly, H. Levine, J. N. Onuchic, and A. Levchenko, *Pericytes Enable Effective Angiogenesis in the Presence of Proinflammatory Signals*, Proc National Acad Sci 116, 23551 (2019).

[15] J. W. Baron and T. Galla, *Intrinsic Noise, Delta-Notch Signalling and Delayed Reactions Promote Sustained, Coherent, Synchronized Oscillations in the Presomitic Mesoderm*, J Roy Soc Interface 16, 20190436 (2019).

[16] M. Boareto, *Patterning via Local Cell-Cell Interactions in Developing Systems*, Dev Biol 460, 77 (2020).

[17] T. Yaron, Y. Cordova, and D. Sprinzak, *Juxtacrine Signaling Is Inherently Noisy*, Biophys J 107, 2417 (2014).

[18] P. S. Swain, M. B. Elowitz, and E. D. Siggia, *Intrinsic and Extrinsic Contributions to Stochasticity in Gene Expression*, Proc National Acad Sci 99, 12795 (2002).

[19] L. S. Tsimring, *Noise in Biology*, Rep Prog Phys 77, 026601 (2014).

[20] A. Singh, *Transient Changes in Intercellular Protein Variability Identify Sources of Noise in Gene Expression*, Biophys J 107, 2214 (2014).

[21] M. B. Elowitz, A. J. Levine, E. D. Siggia, and P. S. Swain, *Stochastic Gene Expression in a Single Cell*, Science 297, 1183 (2002).

[22] M. Lu, J. Onuchic, and E. Ben-Jacob, *Construction of an Effective Landscape for Multistate Genetic Switches*, Phys Rev Lett 113, 078102 (2014).

[23] R. Losick and C. Desplan, *Stochasticity and Cell Fate*, Science 320, 65 (2008).

[24] W. de Back, J. X. Zhou, and L. Brusch, *On the Role of Lateral Stabilization during Early Patterning in the Pancreas*, J Roy Soc Interface 10, 20120766 (2013).

[25] V. Kohar and M. Lu, *Role of Noise and Parametric Variation in the Dynamics of Gene Regulatory Circuits*, Npj Syst Biology Appl 4, 40 (2018).

[26] O. Shaya and D. Sprinzak, *From Notch Signaling to Fine-Grained Patterning: Modeling Meets Experiments*, Curr Opin Genet Dev 21, 732 (2011).

[27] J. J. Hopfield, *Neural Networks and Physical Systems with Emergent Collective Computational Abilities*, Proc National Acad Sci 79, 2554 (1982).

[28] C. B. Anfinsen, *Principles That Govern the Folding of Protein Chains*, Science 181, 223 (1973).

[29] J. Toner and Y. Tu, *Flocks, Herds, and Schools: A Quantitative Theory of Flocking*, Phys Rev E 58, 4828 (1998).

[30] D. U. Ferreiro, A. M. Walczak, E. A. Komives, and P. G. Wolynes, *The Energy Landscapes of Repeat-Containing Proteins: Topology, Cooperativity, and the Folding Funnels of One-Dimensional Architectures*, Plos Comput Biol 4, e1000070 (2008).

[31] F. Font-Clos, S. Zapperi, and C. A. M. L. Porta, *Topography of Epithelial–Mesenchymal Plasticity*, Proc National Acad Sci 115, 201722609 (2018).

[32] G. Toulouse, *Theory of the Frustration Effect in Spin Glasses: I*, Comm. Phys. 2, 115 (1977).

[33] S. F. Edwards and P. W. Anderson, *Theory of Spin Glasses*, J. Phys. F: Metal Phys. 5, 965 (1975).

[34] I. G.-A. Pemartín, V. Martin-Mayor, G. Parisi, and J. J. Ruiz-Lorenzo, *Numerical Study of Barriers and Valleys in the Free-Energy Landscape of Spin Glasses*, J Phys Math Theor 52, 134002 (2019).

[35] R. Benedito and M. Hellström, *Notch as a Hub for Signaling in Angiogenesis*, Exp Cell Res 319, 1281 (2013).

[36] Y. L. Koon, S. Zhang, M. B. Rahmat, C. G. Koh, and K.-H. Chiam, *Enhanced Delta-Notch Lateral Inhibition Model Incorporating Intracellular Notch Heterogeneity and Tension-Dependent Rate of Delta-Notch Binding That Reproduces Sprouting Angiogenesis Patterns*, Sci Rep-Uk 8, 9519 (2018).

[37] R. Goodyear and G. Richardson, *Pattern Formation in the Basilar Papilla: Evidence for Cell Rearrangement*, J Neurosci 17, 6289 (1997).

[38] E. Teomy, D. A. Kessler, and H. Levine, *Ordered Hexagonal Patterns via Notch-Delta Signaling*, Biorxiv 550657 (2019).

[39] J. S. Chen, A. M. Gumbayan, R. W. Zeller, and J. M. Mahaffy, *An Expanded Notch-Delta Model Exhibiting Long-Range Patterning and Incorporating MicroRNA Regulation*, Plos Comput Biol 10, e1003655 (2014).

[40] N. Guisoni, R. Martinez-Corral, J. Garcia-Ojalvo, and J. de Navascués, *Diversity of Fate Outcomes in Cell Pairs under Lateral Inhibition*, Development 144, 1177 (2017).

[41] S. A. V. Mercedes, F. Bocci, H. Levine, J. N. Onuchic, M. K. Jolly, and P. K. Wong, *Decoding Leader Cells in Collective Cancer Invasion*, Nat Rev Cancer 21, 592 (2021).

[42] O. M. J. A. Stassen, T. Ristori, and C. M. Sahlgren, *Notch in Mechanotransduction – from Molecular Mechanosensitivity to Tissue Mechanostasis*, J Cell Sci 133, jcs250738 (2020).

[43] M. Randeria, J. P. Sethna, and R. G. Palmer, *Low-Frequency Relaxation in Ising Spin-Glasses*, Physical Review Letters 54, 1321 (1985).